\documentstyle[titlepage,12pt]{article}
\newcommand{\beq}{\begin{equation}}
\newcommand{\eeq}{\end{equation}}

\title {
\begin{flushright} PSU/TH/147
\end{flushright}
RADIATIVE CORRECTIONS TO P-LEVELS IN THE TWO-BODY QED PROBLEM\\
}

\author{M. I. Eides\thanks{E-mail address:  eides@phys.psu.edu}\\
Department of Physics, Pennsylvania State University,\\
\medskip University Park, Pennsylvania 16802,
USA\thanks{Address up to September 20, 1994} \\
and
\and
Petersburg Nuclear Physics Institute,\\
\medskip
Gatchina, St.Petersburg 188350, Russia\thanks{Permanent address, e-mail:
eides@lnpi.spb.su}\\
\and
I. B. Khriplovich \thanks{E-mail address: khriplovich@inp.nsk.su}
and A. I. Milstein\thanks{E-mail address: milstein@inp.nsk.su}\\
Budker Institute of Nuclear Physics, 630090 Novosibirsk, Russia\\
and Novosibirsk University
}
\date{July 19, 1994}

\begin{document}
\maketitle
\begin{abstract}
The physical origin of the $m\alpha^5$ radiative corrections to $P$-levels
in the two-body QED problem is elucidated. Then we demonstrate that the
next order, $m\alpha^6$, corrections to those levels are due to the anomalous
magnetic moment only.
\end{abstract}

1. The problem of $m\alpha^6$ corrections to $1s~2p~^3P_J$ helium levels
was treated numerically many years ago \cite{sc,dk,ddh,ls}. Corrections of
the same order to positronium $P$-levels were recently calculated
analytically \cite{kmy}. Those corrections are strongly dominated by the
relativistic effects generated by the $(v/c)^4$ expansion terms. The true
radiative corrections are numerically small, being suppressed by the
"geometrical" factor $1/\pi^2$. It was assumed in Refs.\cite{dk,kmy} that
such corrections are due only to the anomalous magnetic moment
contributions.

However, quite recently radiative corrections of a different type were
discussed in relation to the same problems \cite{td}. Though those
contributions are small numerically, the problem is certainly interesting
from a general point of view. Moreover, in helium such a contribution
would be essential for the comparison with experiment.

In the present note we demonstrate accurately that it is indeed
only the anomalous magnetic moment contributions that generate true radiative
corrections of the order $m\alpha^6$ to $P$-states. Unfortunately, paper
\cite{td} is too concise and lacks calculational details; therefore  we
cannot point out the exact cause of disagreement.

\bigskip
2. Let us start with the discussion of the $m\alpha^5$ radiative corrections
to $P$-levels. There are three sources of terms.  First, the infrared
divergence in the scattering amplitude, i.e. in the electron charge radius
(see Fig.1; here and below the dashed line refers to the Coulomb field, the
wavy one to a transverse photon). Being cut off at atomic energies, the
would be divergence  generates both the $\log\alpha$ term and the Bethe
logarithm $L_{n0}$ in the $S$-state Lamb-shift.  For $P$-states the same
divergence leads only to the corresponding Bethe logarithm $L_{n1}$.

The second kind of correction in the two-body problem originates from the
double exchange diagrams presented in Figs. 2-3. Diagrams in Fig.2 with one
Coulomb and one magnetic exchange are also infrared divergent, generating
again $\log\alpha$ and $L_{n0}$ in $S$-states, and $L_{n1}$ in $P$-states
(see, e.g., Ref.\cite{kmy1}). On the other hand, those diagrams are
effectively cut off from above at atomic momenta $q$.  Then, diagrams in
Fig.3 with double magnetic exchange are cut off from below at $q$.  The
$\log q$ contributions from diagrams in Figs.2-3, being transformed into
the coordinate representation, generate a potential $\sim r^{-3}$ which has
nonvanishing (and convergent) expectation value in $P$-states.

And finally, $m\alpha^5$ radiative corrections to $P$-levels are generated
by the $\alpha/2\pi$ contribution to the anomalous magnetic moment.

\bigskip
3. Consider now all potential sources of order $m\alpha^6$ corrections to
the energy shifts of $P$-levels in the same order as we have done above for
the corrections of order $m\alpha^5$.

First of all, the two-loop  contribution to the electron charge radius
is infrared finite. This can be most easily demonstrated using the
Fried-Yennie (FY) gauge for radiative quanta. Therefore there is no
analog of the Bethe logarithm $L_{nl}$ either for $S$- or for $P$-states.

As to the double infrared divergence in the scattering
amplitude connected with emission of two brehmsstrahlung quanta (see Fig.1
for the case of one  brehmsstrahlung quantum), it contains additional power
of momentum transfer squared in comparison with the case of one
brehmsstrahlung photon and is thus capable of producing corrections no
larger than $m\alpha^8$.

Let us pass over now to the two-photon exchange generating the
$m\alpha^5$ approximation $\log q$ contribution to the scattering
amplitude and correspondingly $r^{-3}$ to the interaction potential in
the coordinate representation. We are going to consider radiative
corrections to these graphs and to prove that respective diagrams do not
produce corrections of order $m\alpha^6$ to the energy levels. Note first
that insertion of a one-loop polarization in any of the exchanged photons in
Figs. 2-3 provides an additional factor $q^2$  in the respective integral
for the energy shift. This gets an additional factor $\alpha^2$
after integration.  Hence, polarization  insertions are irrelevant in order
$m\alpha^6$ in the case of exchanged photons with atomic ($\sim
m\alpha$) momenta.

The case of one-loop radiative insertions in the electron lines seems to be
more involved. Once again, we confine for the moment our consideration
to the case of small momenta of exchanged quanta; large ($\sim m\alpha$)
exchanged momenta will be considered below separately. First of all in the
case of two exchanged magnetic quanta the sum of radiative corrections to
the Compton scattering amplitudes entering diagrams in Fig.3  vanish. This
is simply a  direct consequence of the well-known low-energy theorem
\cite{ti} for the Compton scattering.

Consider next radiative photon insertions in the diagrams with one magnetic
and one Coulomb quantum in Fig. 2. FY gauge  for the radiative photons is
again most suitable for our goals since it provides the most smooth low
frequency behavior for all graphs. Let us start with insertion of the
self-energy operator in the electron line in Figs.2. Explicit
expression for the renormalized self-energy operator in  the FY gauge
\cite{tom} (see also \cite{eks}) is proportional to the squared Dirac
operator $(\hat p-m)^2$. Then it is easy to estimate the contribution of
the graphs in Figs.2 with an inserted self-energy operator. The Coulomb line
may be swallowed up by the Schr\"odinger-Coulomb wave function, one of the
Dirac operators in the numerator of the expression for the self-energy is
canceled by the remaining electron propagator and we are left with the
product of the magnetic exchange and the Dirac operator between the
Schr\"odinger-Coulomb wave functions. It is evident that the free Dirac
operator $\hat p-m$ applied to the Schr\"odinger-Coulomb wave function then
produces a factor $\alpha^2$. Therefore, self-energy insertion in the
electron line suppresses the previous order effect not by a factor of
$\alpha$ but by $\alpha^3$.

Consider now radiative photon insertion in one of the vertices in Fig.2.
Respective anomalous magnetic moment contribution produces correction of
order $m\alpha^6$ as was mentioned above. All other terms in the one-loop
vertex correction in the FY gauge contain at least one additional
suppression factor (see \cite{eks}) which after loop integration leads to
contribution to the energy shift of order $m\alpha^7$. We also have to
consider insertion of a spanning radiative photon. Once again, as was
shown in \cite{eks}, respective diagrams contain in the FY gauge
an additional suppression factor which turns into an additional factor
$\alpha^2$, leading to the contribution of order $m\alpha ^7$ which is too
small to be of interest for us now.

As to the diagrams with multiple exchange by soft ($q\sim m\alpha$) Coulomb
quanta, insertion of a radiative spanning photon leads only  to the
$m\alpha^5$ Bethe logarithm $L_{n1}$.

Consider finally graphs with two exchanged photons of high ($\sim m$)
momenta. It is easy to see that in these diagrams all one-loop radiative
insertions either in the electron, or in the exchange photon line produce
corrections of order $m\alpha^6$ only to $S$-states.

Up to this moment we deliberately omitted graphs with the anomalous
magnetic moment corrections. These graphs produce indeed the $m\alpha^6$
corrections to the $P$-levels. There are no other sources of corrections of
this order.

\bigskip

The authors are deeply grateful to H.~Grotch, S.~G.~Kar\-shen\-boim,
V.~A.~She\-lyu\-to and A.~S.~Yelkhovsky for use\-ful dis\-cus\-sions.
M.~I.~Ei\-des and I.~B.~Khrip\-lo\-vich are extremely grateful to H.~Grotch
for the warm hospitality extended to them at Penn State University where
part of this work has been done. Work of M.~I.~Eides was supported in part
by Grant \#R2E000 from the International Science Foundation and by Grant
\#02-3853 of the Russian Foundation for Fundamental Research.
I.~B.~Khriplovich and A.~I.~Milstein acknowledge the financial support by
the program "Universities of Russia", Grant \#94-6.7-2053.

\newpage
\begin{center}\large Figure Captions
\end{center}
\vskip5cm
\noindent
Fig.1. Diagrams with infrared divergent one-loop charge radius insertion.\\

\noindent
Fig.2 Diagrams with one Coulomb and one magnetic exchange.\\

\noindent
Fig.3 Diagrams with double magnetic exchange.

\end{document}